\documentclass[aps,a4paper,twocolumn,floatfix,showpacs,
groupedaddress]{revtex4}

\usepackage{bm,amsmath,amssymb,amsfonts}
\usepackage{array}
\usepackage[dvips]{graphicx}

\usepackage{color}
\definecolor{dred}{rgb}{0.6,0.0,0.0}

\begin{document}

\title{\textcolor{dred}{Engineering  light localization in a fractal waveguide network}}

\author{Biplab Pal}
\author{Pinaki Patra} 
\author{Jyoti Prasad Saha}
\author{Arunava Chakrabarti}
\email[\textbf{E-mail:} arunava\_chakrabarti@yahoo.co.in]{}
\affiliation{Department of Physics, University of Kalyani, Kalyani,
West Bengal-741 235, India}

\begin{abstract}
We present an exact analytical method of engineering the localization of 
electromagnetic waves in a fractal waveguide network. It is shown that, 
a countable infinity of localized electromagnetic modes with a multitude of 
localization lengths can exist in a Vicsek fractal geometry 
built with diamond shaped monomode waveguides  as the `unit cells'. 
The family of localized modes form clusters of increasing size. 
The length scale at which the onset of localization for each mode 
takes place can be engineered at will, following a well defined prescription 
developed within the framework of a real space renormalization group. 
The scheme leads to an exact evaluation of the wave vector for every such 
localized state, a task that is non-trivial, if not impossible for any random 
or deterministically disordered waveguide network.   
\end{abstract}

\maketitle

%%%%%%%%%%%%%%%%%%%%%%%%%%%%%%%%%%%%%%%%%%%%%%%%%%%%%%%%%%%%%%%%%%%%%%%%%%%%%%%%%%
\section{Introduction}
Since its inception, the phenomenon of Anderson localization  
in a disordered lattice~\cite{pwa1} has been found to be ubiquitous in 
diverse fields of condensed matter physics and materials science. While the 
popular domain of interest is related to electronic systems, where the 
quantum interference plays a pivotal role in localizing electronic 
eigenstates in presence of 
disorder~\cite{angus,thouless,schreiber,alberto1,alberto2,zilly}, the effect 
is by no means, confined to electrons, and extend over a variety of 
phenomena ranging from spin freezing in one dimensional 
semiconductors~\cite{echeverria}, localization in optical 
lattices~\cite{sankar1,edwards,roati}, or the localization of matter waves 
(cold atoms forming Bose-Einstein condensates)~\cite{gavish,lye}, to name a 
few. Incidentally, the latter has recently been observed experimentally 
in one dimensional matter waveguides, where the random potential is 
generated by laser speckles~\cite{billy}. 

In 1984 Sajeev John pointed out that the idea of localization goes far 
beyond the electronic systems, and is actually a general phenomenon common 
to any wave propagation in systems with disorder~\cite{john}. Anderson 
followed with a seminal paper considering the idea of localization of 
classical waves, in an attempt to work out the theory of white 
paints~\cite{pwa2}. The field gathered momentum in the last couple of 
decades and a considerable volume of literature related to the localization 
of classical waves, particularly light, ultrasound and microwave is now in 
existence~\cite{jovic1,bliokh,jovic2,sutherland,asatryan,hodges,he,yablonovich1,shi}. 

The study of localization of light in disordered media has been patronized 
by the discovery of the photonic band gap (PBG) 
materials~\cite{soukoulis,john2,yablonovitch2}. These systems exhibit gaps 
in the frequency spectrum in which propagation of electromagnetic waves is 
forbidden. This has important implications in both fundamental science and 
technological applications.

Photonic gaps, apart from materials with a large dielectric constant, can 
also be observed in waveguide networks, as proposed by several groups over 
the past years~\cite{zhang,dobrzynski,vasseur,pradhan,li,sheelan1,sheelan2,lu1,lu2}. 
Anderson localized eigenmodes are observed inside the photonic gaps and 
excellent agreement between theory and experiments has been 
obtained~\cite{zhang}. The network models are able to localize a propagating 
wave by virtue of the geometrical arrangement of the waveguide segments. Of 
particular interest are a wide variety of models based on waveguide networks 
designed following a deterministic fractal 
geometry~\cite{li,sheelan1,sheelan2,lu1,lu2}, where the gaps result from the 
typical topology exhibited the hierarchical arrangement of the 
waveguide segments. The present communication also deals with a 
hierarchically designed (fractal geometry) waveguide network, but addresses 
a deeper fundamental question regarding wave localization in such systems, 
as explained below.

Fractals or hierarchical geometries in general cause an excitation to localize~\cite{kadanoff,rammal,kappertz,schwalm1}. The energy spectrum 
turns out to be singular continuous~\cite{kadanoff}, with a gap in the 
vicinity of every energy. Nevertheless, there can be a countable 
infinity of {\it extended} eigenfunctions with high (or even, perfect) 
transmittivity, even though there is no translational order in such systems. 
Once again, this is true for electrons~\cite{bibhas,anirban,xiang,arun1,macia,moritz,arun2}, 
and classical waves as well~\cite{sheelan1,sheelan2}. A curious point, 
apparently gone unnoticed or un-appreciated so far is that, while a precise 
determination of the eigenvalues corresponding to the extended eigenmodes is 
possible in the above cases of hierarchically grown fractal networks, the 
task seems to be practically impossible when it comes to an exact evaluation 
of eigenvalues of the localized modes in such  hierarchical systems in their 
{\it thermodynamic limits}. Direct diagonalization of the Hamiltonian (in 
the electronic case), or an exact numerical solution of the wave equation 
doesn't help, as the overall character of the spectrum in all the cases is 
highly fragmented, and the eigenvalues obtained from a finite sized network 
are likely to {\it slip away} from the spectrum once we go over to a higher 
generation. This problem has recently been addressed in the context of 
electron localization in fractal space~\cite{biplab}, and we carry forward 
the central idea floated in Ref.~\cite{biplab} to evaluate the {\it exact} 
wavelengths (wave vectors) of electromagnetic waves that can be localized 
{\it at will} in a properly designed hierarchical single channel network. To 
the best of our knowledge, this issue remains unaddressed so 
far in the literature in the context of localization of wave, light in particular. 

\begin{figure*}[ht]
\includegraphics[clip,width=16cm]{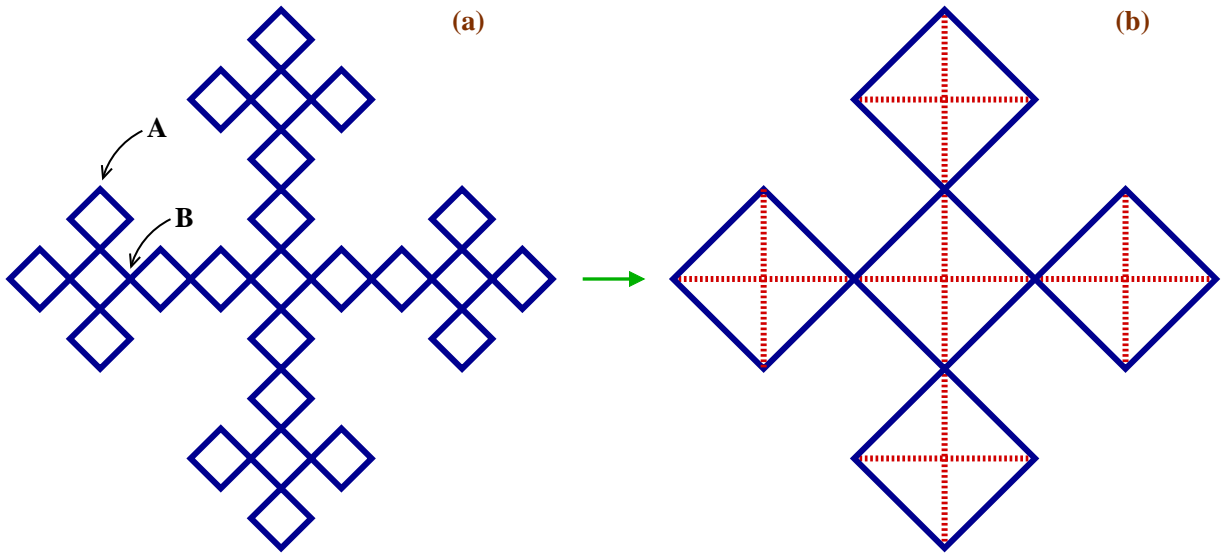}
\caption{ (Color online) (a) Schematic view of the second generation of an 
infinite diamond-Vicsek waveguide network. The vertices at the junction of 
two waveguides are named as `$A$' in the text, while those in the bulk (at 
the crossing of four waveguides) are termed `$B$'. (b) Renormalized version 
of (a) with the dotted lines indicating the diagonal `hopping' which is 
generated due to renormalization.}
\label{system}
\end{figure*}

We design a Vicsek fractal network~\cite{vicsek} consisting of {\it diamond} 
shaped loops each arm of which mimics a single mode linear waveguide~[see 
Fig.~\ref{system}]. While examination of the localized mode  eigenvalues and 
the nature of localization are indeed the major factors motivating this work, 
other interests in such a study are related to the general spectral 
character and classical wave transport in these systems.
%######################################################
%######################################################
As one can easily appreciate, the geometry of a diamond-Vicsek network 
provides an interesting culmination of the `open' character of a typical 
Vicsek pattern and closed loops at shorter scales of length. This is in 
marked contrast to the much studied Sierpinski gasket waveguide 
network~\cite{li}, which is a closed structure, or to the other 
deterministic waveguide networks~\cite{sheelan2,lu1}. The presence of the 
loops generates a possibility of an effectively long ranged propagation of waves 
between the various vertices, and its effect on the localization or de-
localization of waves is worth studying. 

We find interesting results. For an infinite hierarchical 
geometry, such as presented above, a countable infinity of eigenmodes with a 
multitude of localization lengths can be {\it precisely detected}. One can 
work out an exact mathematical prescription to specify the length scale at 
which the onset of localization begins. The localization can in principle, 
be {\it delayed} (staggered) in position space and  the corresponding wave 
vectors (or, wavelengths) can be exactly evaluated following the same 
prescription based on a real space renormalization group (RSRG) method~\cite{southern}. 
In addition, it is shown that for a given set of parameters, the center of the 
spectrum corresponds to a perfectly extended eigenmode, with the parameters 
describing the system exhibiting a fixed point behavior. This central 
extended mode is flanked on either side by localized wave functions with a 
hierarchy of localization lengths.

In what follows we describe the results. In Sec.~\ref{sec2}, the model and 
the mathematical method of handling the problem are presented. 
Section~\ref{sec3} discusses the results and their analyses, and in 
Sec.~\ref{sec4} we draw our conclusions . 
%%%%%%%%%%%%%%%%%%%%%%%%%%%%%%%%%%%%%%%%%%%%%%%%%%%%%%%%%%%%%%%%%%%%%%%%%%%%%%%%%%
\section{The model and the method}
\label{sec2}
%%%%%%%%%%%%%%%%%%%%%%%%%%%%%%%%%%%%%%%%%%%%%%%%%%%%%%%
\subsection{The wave equation and its discretization}
We have considered a waveguide network formed by waveguide segments having 
the same lengths arranged in a Vicsek fractal geometry~\cite{vicsek}. Each 
segment has a single channel for wave propagation. The wave function 
$\psi_{ij}$ between any two nodal points $i$ and $j$ satisfies the wave 
equation:
\begin{equation}
\label{waveeqn}
\frac{\partial^2 \psi_{ij}(x)}{\partial x^2}
+\frac{\omega^2}{c^2}\psi_{ij}(x) = 0
\end{equation}
where $\omega$ is the frequency of the wave, $c$ is the speed of wave 
propagation and $x$ is the distance measured from the $i$th node. The above 
equation has the solution of the form~\cite{alexander,ping}
\begin{equation}
\label{soln}
\psi_{ij}(x)= \psi_{i}\frac{\sin[k(\ell_{ij}-x)]}{\sin(k\ell_{ij})} 
+ \psi_{j}\frac{\sin(kx)}{\sin(k\ell_{ij})}
\end{equation}
where $k=\omega/c$, $\ell_{ij}$ is the length of the segment between the 
node $i$ and $j$, and $\psi_{i}$ and $\psi_{j}$ are the   values of the wave 
function at the $i$th and $j$th nodes respectively. The flux conservation 
condition:
\begin{equation}
\label{condition}
\sum_{j}\left[\frac{\partial}{\partial x}\psi_{ij}(x)\right]_{x=0} = 0
\end{equation}
where the summation $j$ is over all the nodes linked directly to $i$, leads to a discretized version of Eq.~\eqref{waveeqn}~\cite{ping}, viz.,
\begin{equation}
\label{difference1}
-\psi_{i}\sum_{j}\cot\theta_{ij}+\sum_{j}\psi_{j}/\sin\theta_{ij} 
= 0
\end{equation}
where $\theta_{ij}= k\ell_{ij} = ka$, $a$ being the constant length of a 
waveguide, as considered in all the calculations which follow. 
Eq.~\eqref{difference1} resembles a tight binding difference equation 
depicting the propagation of non-interacting electrons in a lattice, viz.,
\begin{equation}
\label{difference2}
(E-\epsilon_{i})\psi_{i}=\sum_{j}t_{ij}\psi_{j}
\end{equation}
with the `electron energy' $E$ can be put equal to $2\cos ka$, the `on-site 
potential' $\epsilon_{i}=2\cos ka+\sum\limits_{j}\cot \theta_{ij}$ and 
`hopping matrix element' $t_{ij}=1/\sin\theta_{ij}$. It should be 
appreciated that the choice of $E = 2 \cos ka$ is completely arbitrary, and 
this does not affect the final results in any way. As seen in 
Fig.~\ref{system}, the Vicsek waveguide network has two types of nodal 
points, viz., type $A$ (having two neighboring nodal points) and type $B$ 
(having four neighboring nodal points). Accordingly, the equivalent `on-site 
potentials' are assigned two values, viz.,  
$\epsilon_{A}= 2\cos ka+2\cot ka$ and   
$\epsilon_{B}= 2\cos ka+4\cot ka$ respectively. The `overlap integral' along 
an arm of the waveguide is $t = 1/\sin ka$. There is no second neighbor 
tunneling of the wave to begin with. But it will grow with RSRG. 

We now proceed to describe the physics of wave propagation in such a fractal 
geometry by exploiting this exact analogy with the corresponding electronic 
problem.
%%%%%%%%%%%%%%%%%%%%%%%%%%%%%%%%%%%%%%%%%%%%%%%%%%%%%%%
\subsection{The RSRG scheme}
A renormalized version of the fractal network is easily obtained by 
decimating a subset of vertices from the original geometry. This implies one 
has to eliminate a subset of the wave amplitudes from the difference 
equations~\eqref{difference2} in terms of the surviving vertices [see 
Fig.~\ref{system}(b)]. This results in the following set of recursion 
relations for the system parameters.
%%%%% RECURSION RELATIONS %%%%%%
\begin{widetext}
\begin{eqnarray}
\epsilon_{A}^{\prime} &=& \epsilon_{A} +\left[\frac{2t^2}{E-\epsilon_{A}- \lambda} + 
\frac{\left(\lambda + \dfrac{2t^2}{E-\epsilon_{A}- \lambda}\right)^{2}
\left\{(E-\epsilon_{B})-\dfrac{2t^2}{E-\epsilon_{A}- \lambda}-\xi\right\}}
{\left\{(E-\epsilon_{B})-\dfrac{2t^2}{E-\epsilon_{A}- \lambda}-\xi\right\}^{2}
-\left(\lambda + \xi\right)^{2}}\right],\nonumber\\
\epsilon_{B}^{\prime} &=& \epsilon_{B} +2\left[\frac{2t^2}{E-\epsilon_{A}- \lambda} + 
\frac{\left(\lambda + \dfrac{2t^2}{E-\epsilon_{A}- \lambda}\right)^{2}
\left\{(E-\epsilon_{B})-\dfrac{2t^2}{E-\epsilon_{A}- \lambda}-\xi\right\}}
{\left\{(E-\epsilon_{B})-\dfrac{2t^2}{E-\epsilon_{A}- \lambda}-\xi\right\}^{2}
-\left(\lambda + \xi\right)^{2}}\right],\nonumber\\
t^{\prime} &=& \frac{t\left\{\lambda(E-\epsilon_{A}- \lambda)+2t^2\right\}^{2}}
{\left\{E^2-(\epsilon_{A}+\epsilon_{B})(E-\lambda)
-2E\lambda +\lambda^{2}-2t^2+\epsilon_{A}\epsilon_{B}\right\}^{2}
-4t^2(E-\epsilon_{A}-\lambda)^{2}},\nonumber\\
\lambda^{\prime} &=& \frac{\left(\lambda + \dfrac{2t^2}{E-\epsilon_{A}- \lambda}\right)^{2}
\left(\lambda + \xi\right)}
{\left\{(E-\epsilon_{B})-\dfrac{2t^2}{E-\epsilon_{A}- \lambda}-\xi\right\}^{2}
-\left(\lambda + \xi\right)^{2}}.
\end{eqnarray}
\end{widetext}
where $\xi=2t^2(E-\epsilon_{A}-\lambda)/\Delta$ with 
$\Delta=(E-\epsilon_{A}-\lambda)(E-\epsilon_{B}-\lambda)-2t^{2}$, 
and $\lambda=0$ at the beginning.

The scaling generates an effective second neighbor `hopping' (overlap) 
$\lambda^{\prime}$, as is obvious from the above set of recursion relations 
and the diagram~\ref{system}(b). It is important to appreciate that, the 
growth of a $\lambda^{\prime}$, unlike the 
electronic case~\cite{biplab}, does not have 
any physical significance here, as there is no real `overlap' of the wave amplitudes.
The light is essentially confined within the waveguide.

These recursion relations will now be used to obtain information about the 
local density of eigenmodes at specific sites of the system, and the 
localized or extended character of the modes, as discussed below.
%%%%%%%%%%%%%%%%%%%%%%%%%%%%%%%%%%%%%%%%%%%%%%%%%%%%%%%%%%%%%%%%%%%%%%%%%%%%%%%%%%
\section{Results and discussion}
\label{sec3}
\subsection{Local density of eigenmodes}
\begin{figure}[ht]
\includegraphics[clip,width=7cm,angle=-90]{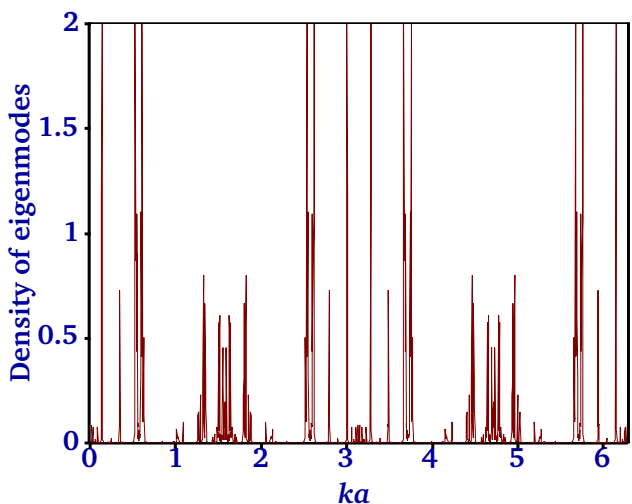}
\caption{ (Color online) Local density of eigenmodes at a bulk $B$-type 
vertex of an infinite diamond-Vicsek waveguide network.}  
\label{ldos1}
\end{figure}

As already stated, the exact mapping of the wave equation on to a discrete 
Schr\"{o}dinger type equation allows us to extract information about the 
density of electromagnetic modes through a Green's function 
analysis~\cite{southern}. We present in Fig.~\ref{ldos1} the density of 
modes at a $B$-vertex, which is given by,  
\begin{equation}
\rho^{(B)}(ka)= \lim_{\eta \rightarrow 0}\,\left[-\dfrac{1}{\pi}\,
\text{Im}\,\{G^{(B)}(ka + i\eta)\}\right]
\label{dos}
\end{equation}
The distribution of eigenmodes, plotted within $ka < 0 < 2\pi$, shows 
clusters of non-zero values over a finite range of the wave vector $k$, and 
is found to be symmetric around $ka = \pi/2$. The fragmented Cantor-like 
character, typical signature of a fractal spectrum, is apparent. An idea 
about the character of the eigenmodes can be obtained by observing the flow 
of $t$ under successive RSRG iterations. In general, for an arbitrary value 
of $ka$, for which the density of modes is non-zero, $t^{(n)} \rightarrow 0$ 
as the number of iterations $n$ increases, implying a localized character of 
the corresponding eigenmode~\cite{southern,bibhas}. However, for $ka = (2m+1) \pi/2$, 
both the first and the second neighbor `hopping integrals' $t$ and $\lambda$ 
remain non-zero for an indefinite number of iterations. In fact, we observe 
a one cycle {\it fixed point} of the entire parameter space, viz., 
$\{\epsilon_{A}^\prime, \epsilon_{B}^\prime, t^\prime, \lambda^\prime\} = 
\{\epsilon_{A}, \epsilon_{B}, t, \lambda\}$.
The fact that $t$ and $\lambda$ remain finite at all stages of RSRG implies 
that there is a non-zero overlap between the wave amplitudes at all scales 
of length, and the corresponding mode is an extended one.

The neighborhood of $ka= (2m+1)\pi/2$ has been scanned minutely. The self 
similarity of the spectrum is always seen with dense patches of eigenvalues 
clustered throughout the intervals. For many of these eigenvalues $t$ and 
$\lambda$ remain finite under successive decimation for a large number of steps. 
This indicates that every local {\it band center} at $ka = (2m+1) 
\pi/2$ is flanked either by extended modes or at least, eigenmodes with very 
large localization lengths. 
%######################################################
%######################################################
%%%%%%%%%%%%%%%%%%%%%%%%%%%%%%%%%%%%%
\subsection{Explicit construction of localized modes}
%%%%%%%%%%%%%%%%%%%%%%%%%%%%%%%%%%%%%
The deterministic Vicsek fractal waveguide is self-similar at all scales of 
length. This feature allows us to explicitly construct a special 
distribution of wave functions by suitably exploiting the difference 
equations Eq.~\eqref{difference2}. These `special' eigenmodes are localized 
and extend over clusters of single channel waveguide segments of various 
sizes. The planar extent of such clusters depends on the eigenvalue 
corresponding to the localized mode, and can be small or enormous, depending 
on the wavelength (or, wave vector). The construction is similar to our 
recent work~\cite{biplab} on electronic states.

\begin{figure*}[ht]
\includegraphics[clip,width=16cm]{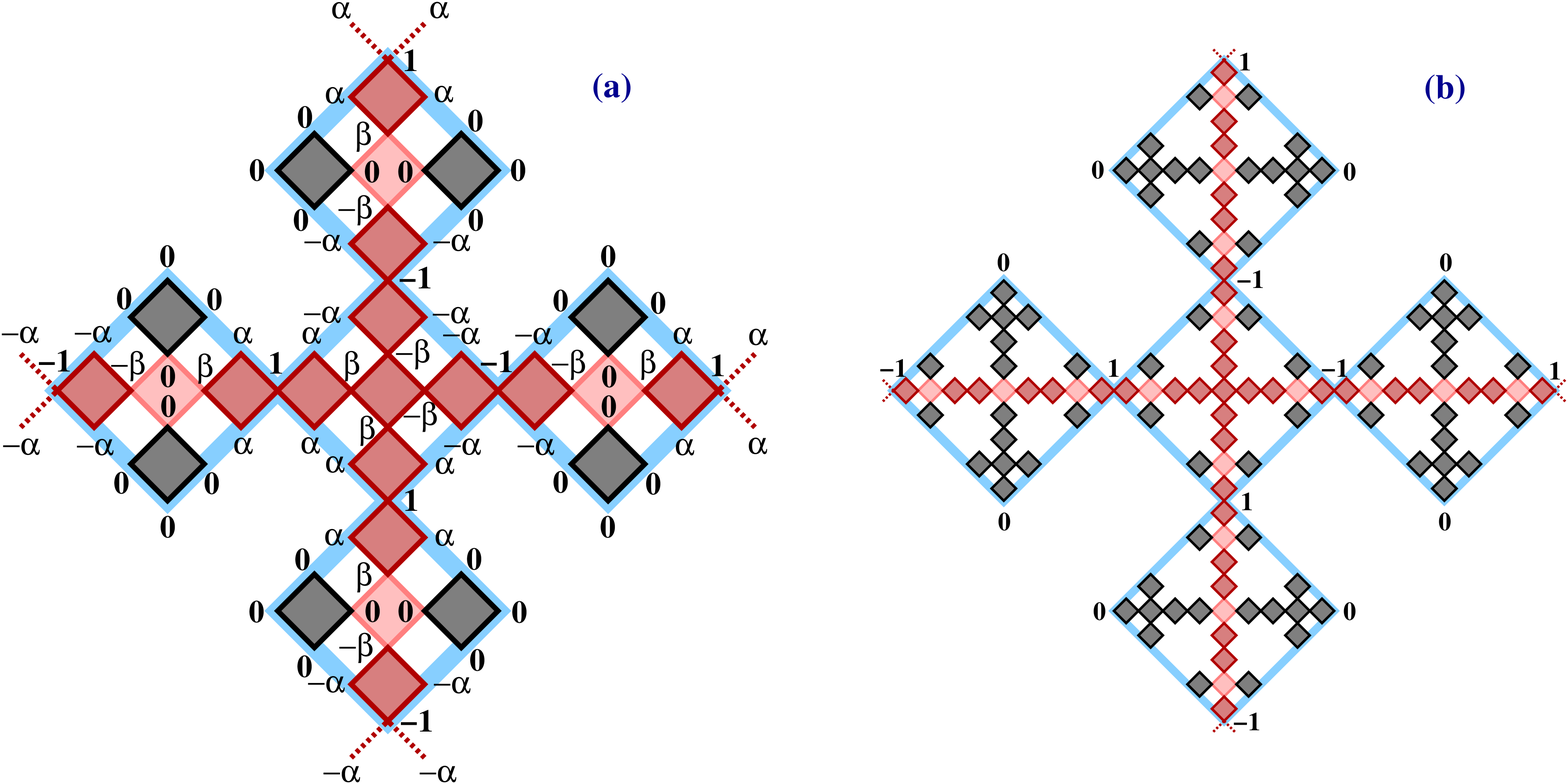}
\caption{ (Color online) (a) Distribution of amplitudes of the wave 
function for $ka = \pi/6$ (obtained by solving Eq.~\eqref{roots} for $n=1$) 
on a second generation network. The dark shaded plaquettes with black lines 
at the boundary embrace network vertices with zero amplitude. 
The deep red waveguide segments (covering a dark red shaded region) connect vertices with 
non-zero amplitude, and `glow' with maximum intensity. The lighter red lines 
represent waveguide segments which have an intermediate intensity profile.
The amplitudes of the wave are marked by the numbers $\pm 1$, $0$, $\pm \alpha$ and 
$\pm \beta$ respectively, where $\alpha=-\sqrt{3}/2$ and $\beta=1/2$.
(b) The distribution of wave amplitudes for $ka = \pi/6$ on a third generation 
network. The thick blue highlighted lines represent one- and two-step 
renormalized lattices in (a) and (b) respectively.}
\label{ampdistribution1}
\end{figure*}

To elaborate, let us set 
\begin{equation}
E = \epsilon_B(n) - 2 \lambda(n)
\label{roots}
\end{equation}
where, $n$ refers to the stage of renormalization. This is in general, a 
polynomial equation in $E$ (and hence, in $k$). The zeros of this equation 
will be the allowed wave vectors for the infinite system if, and only if, 
with them, one can satisfy Eq.~\eqref{difference2} locally at every vertex 
of the network. This task can be accomplished by trying to draw a non-
trivial distribution of amplitudes for a value of $E$ ($ka$) obtained from 
Eq.~\eqref{roots} on the undecimated vertices of an $n$-step renormalized 
network, and then trying to figure out the amplitude distribution on the 
original waveguide structure at the bare length scale. This can indeed be 
done, as we demonstrate in Fig.~\ref{ampdistribution1}(a) for $n=1$. 
%########################################################
%######################################################## 
For $n=1$ Eq.~\eqref{roots} reduces to 
\begin{equation}
\cos ka(2\cos 2ka-1)=0
\end{equation}
Roots of the above equation are $ka = \pm \pi/2$ and $ka = \pm \pi/6$. $ka = \pi/2$ (or, 
equivalently, $-\pi/2$) is of course, the extended mode. The root $ka = \pi/6$ 
leads to the construction of wave amplitudes as shown in Fig.~\ref{ampdistribution1}. 
It is not difficult to extend the construction depicted in 
Fig.~\ref{ampdistribution1}(a) even to a network of an arbitrarily large 
size, where the {\it end} vertices are not actually visible. We are still 
able to satisfy Eq.~\eqref{difference2} locally at every vertex while 
drawing this distribution and thus,  $k =\pi/6a$ definitely belongs to 
the spectrum of the infinite system, a fact that has been cross-checked by 
evaluation the LDOS at the $A$- and the $B$-sites at this special value of 
$k$. We get a stable, finite value of the LDOS which supports our argument 
above.

In Fig.~\ref{ampdistribution1}(a) we show the distribution of amplitudes on 
the central cluster of an infinite diamond-Vicsek hierarchical network for 
$ka =\pi/6$. The deep red arms  connect network vertices where the wave 
amplitudes are non-zero, and thus these arms are the brightest looking ones 
as far as the distribution of light intensity is concerned. The black lines represent 
waveguides which will appear completely dark, as the wave amplitudes at 
their vertices will have to be zero in order to satisfy Eq.
\eqref{difference2}. There will be arms connecting one vertex with zero 
amplitude and another with a non-zero one. These are depicted by a lighter shed of 
red, and will `glow' with less intensity compared to the deep red ones. The 
distribution of intensity in any arm (apart from the black ones) is by no 
means uniform. The colors just represent the fact that the intensity is non-
zero. The significant observation is that, clusters of non-zero amplitude 
span over a finite distance, but ultimately get `decoupled' from each other 
on a larger scale of length. This can be appreciated if we look at 
Fig.~\ref{ampdistribution1}(b) which is a larger version of the previous 
figure. The red shaded clusters are distributed along the principal $X$- and 
$Y$-axes, but are {\it separated} from each other beyond a certain extent by 
light red boxes. The black clusters representing amplitude-voids are now 
seen to span larger spatial  distances. A similar construction is possible 
for $ka =-\pi/6$ which is another solution of Eq.~\eqref{roots} for $n=1$, 
but this does not have any additional significance. 
In terms of light, the entire hierarchical geometry will have an appearance 
where light will be {\it localized} with higher intensity at certain 
clusters of waveguides, decoupled from each other by completely dark 
patches.

It is apparent from the above discussion that the eigenfunction 
corresponding to $ka = \pm \pi/6$ will be localized in the fractal space. 
This is easily re-confirmed by studying the evolution of the hopping 
integrals under successive RSRG steps. The hopping integrals $t$ and 
$\lambda$ (zero initially, but grows later) remain non-zero at the first 
stage of RSRG (that is, $n=1$), indicating that the nearest neighboring 
sites on a one step renormalized lattice will have a non-zero overlap of the 
wave functions. They start decaying for $n>2$ with the decay in $\lambda(n)$ 
taking place at a much slower rate compared to $t(n)$. This indicates that 
over larger scale of length the corresponding states are {\it localized}, 
but the effect is a weak one.
%%%%%%%%%%%%%%%%%%%%%%%%%%%%%%%%%%%%%
\subsection{The staggering effect}
%%%%%%%%%%%%%%%%%%%%%%%%%%%%%%%%%%%%%
The previous observation immediately leads to an innovative way of exactly 
determining the wave vectors (wavelengths) corresponding to localized wave 
functions on such a deterministic geometry at an arbitrary scale of length. 
We do it using the following method.

We can solve Eq.~\eqref{roots} in principle, to get the desired $k$-values 
for any $n$. For example, we have done it explicitly for $n=2$. The roots 
are obtained from the equation,
\begin{equation}
\cos ka(2\cos 2ka-1)(\cos 2ka-2\cos 8ka)=0
\end{equation}
and are given in Table~\ref{table}.
%#######################
\begin{table}
\begin{tabular}{|>{\centering\arraybackslash}m{0.4in}|>{\centering\arraybackslash}m{1.4in}|}
\hline
{\boldmath\(n\)} & {\bf Values of \boldmath\(ka\)}\\
\hline
$1$ & $\pm 1.570796$, $\pm 0.523599$\\
\hline
& $\pm 1.570796$, $\pm 0.523599$, \\
& $\pm 2.617994$, $\pm 1.823731$, \\
$2$ & $\pm 1.317862$, $\pm 2.132603$, \\
& $\pm 1.008990$, $\pm 2.531027$, \\
& $\pm 0.610566$, $\pm 3.008153$, $\pm 0.133440$ \\
\hline
\end{tabular}
\caption{Values of $ka$ obtained from the Eq.~\eqref{roots} for $n=1$ and 
$n=2$ respectively.}
\label{table}
\end{table}
%#######################
In every case, on beginning the RSRG iteration with the wave vector $k$ 
chosen arbitrarily from the above set, the nearest neighbor overlap integral 
$t$, and the diagonal one, viz., $\lambda$ remain non zero {\it at least up 
to} that specific $n$-th stage of renormalization. After that, as the RSRG 
progress, the hoppings flow to zero with $\lambda$ dominating over $t$ at 
every step of renormalization. This implies that for any such $k$-value one 
can draw a non-trivial distribution of wave-amplitudes on the renormalized 
fractal network. When mapped back on to the original lattice the amplitudes 
will be found to span clusters of increasing size. The exact size of the 
spanning clusters will be determined  by the value of $n$. The size, for 
example, with $n=2$ exceeds that for $n=1$.

The spanning clusters finally get decoupled from similar clusters when one 
looks at the distribution over a large enough network. Speaking in terms of 
the red and grey shaded zones, it should be appreciated that the size of the 
red zone is much bigger for $n=2$ compared to the $n=1$ case. We refer the 
reader to Ref.~\cite{biplab} for understanding the result.

It is now obvious that higher the value of $n$, more will be the number of roots 
of the polynomial equation Eq.~\eqref{roots}. The roots have a nice nesting 
property. The roots obtained from any $(n-1)$-th stage 
are found included in the solution for the $n$-th stage (see Table~\ref{table}). 
The additional 
roots obtained at the $n$-th level over the existing roots from the $(n-1)$-
th level keep the overlap integrals non-zero up to that specific $n$-th RSRG 
step. Beyond this step, the overlap finally starts to weaken in magnitude. 
This immediately implies that for larger values of $n$, the clusters of non-
zero amplitudes span wider fractal space, and the localization effect begins 
much later. That is, on-set of localization can be {\it delayed in space}, at 
one's will, by choosing to solve Eq.~\eqref{roots} for larger value of $n$. 
We can thus bring a staggering effect on the localization of light in such a 
fractal waveguide network.
%%%%%%%%%%%%%%%%%%%%%%%%%%%%%%%%%%%%%%%%%%
\subsection{Transmission of electromagnetic waves}
%%%%%%%%%%%%%%%%%%%%%%%%%%%%%%%%%%%%%%%%%%
To get the two terminal conductance for a finite size diamond-Vicsek 
fractal, we attach the system between two semi-infinite one-dimensional 
single channel waveguides. The wave equation obeyed by the incident wave in 
these waveguides is discretized, and the leads are artificially converted 
in to arrays of effective {\it nodes} characterized by a constant on-site 
potential $\epsilon_{l} = 2 \cos{ka} + 2 \cot{ka}$ as before,  and a nearest 
neighbor overlap integral $t_{l} = 1/\sin{ka}$. We then successively renormalize 
the finite network to reduce it into an effective two-vertex system, with 
renormalized effective on-site term equal to $\mathcal{U}$ and with 
an effective hopping integral $\mathcal{T}$ between them. The transmission 
coefficient across the effective dimer is given by the well known 
formula~\cite{stone},
\begin{equation}
T=\dfrac{4\sin^{2}ka}{\mathcal{D}_{1}^{2}+\mathcal{D}_{2}^{2}}
\end{equation}
where $\mathcal{D}_{1}=\left[(M_{12}-M_{21})+(M_{11}-M_{22})\cos ka 
\right]$ and 
$\mathcal{D}_{2}=\left[(M_{11}+M_{22})\sin ka \right]$. 
The matrix elements $M_{ij}$ are given by, 
$M_{11} =\dfrac{(E-\mathcal{U})^{2}}{\mathcal{T}t_{l}}
-\dfrac{\mathcal{T}}{t_{l}},\ 
M_{12} =-\dfrac{(E-\mathcal{U})}{\mathcal{T}},\ 
M_{21} =-M_{12},\ 
M_{22} =-\dfrac{t_{l}}{\mathcal{T}}$
and $\cos ka = (E-\epsilon_{l})/2t_{l}$, `$a$' being the lattice constant 
and is taken to be equal to unity throughout the calculation.
%##########################################################
\begin{figure}[ht]
\centering 
\includegraphics[clip,width=7cm,angle=-90]{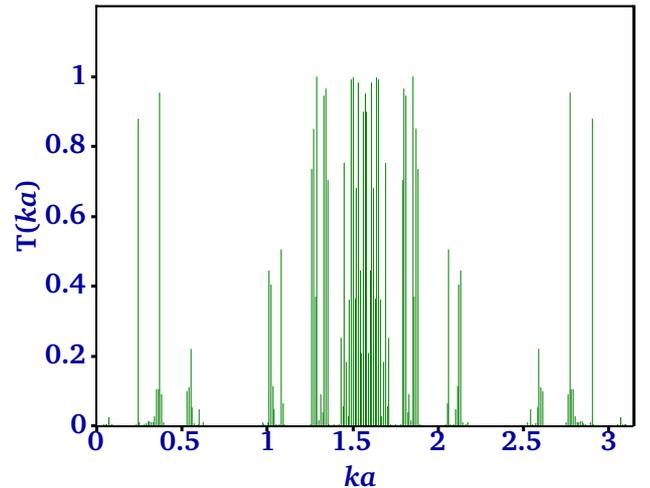}
\caption{ (Color online) Transmission characteristics for a $3$rd generation 
Vicsek waveguide network. The length of each monomode segment $a$ is chosen 
as unity.} 
\label{trans}
\end{figure}
%########################################################## 

In Fig.~\ref{trans}, we have shown the two-terminal transmission 
characteristics for a 3rd generation waveguide network system. The 
transmission spectrum is, as expected, full of gaps, with resonant 
transmission exhibited at certain $k$-values. With increasing generation, 
the resonances become rarer and the spectrum become much more fragmented.
%%%%%%%%%%%%%%%%%%%%%%%%%%%%%%%%%%%%%%%%%%%%%%%%%%%%%%%%%%%%%%%%%%%%%%%%%%%%%%%%%%   
\section{Concluding remarks}
\label{sec4}
In conclusion, we have examined the distribution of intensity of light (or 
any electromagnetic wave) on a Vicsek geometry consisting of diamond shaped 
single channel waveguides. The major result is that we have been able to 
identify a countable infinity of localized eigenmodes displaying a multitude 
of localization lengths. A prescription is given for an exact determination 
of the wave vectors corresponding to all such modes, a problem that is far 
from trivial in the case of a deterministically disordered system. The 
localized wave functions span the fractal space in clusters of 
increasing sizes, the size being precisely controlled by the length scale at 
which the wave vector ($k$) is evaluated. The onset of localization can be 
exactly predicted from the stage of RSRG and can be delayed (staggered) in 
space. The study provides a unique 
opportunity to experimentally observe the localization of light triggered by 
the lattice topology without bothering about the high dielectric constant 
materials. It may also be useful in developing novel photonic band gap 
structures, where the wavelengths to be screened or allowed to go through 
can be controlled over arbitrarily small domains exploiting the fractal 
character of the network.
%%%%%%%%%%%%%%%%%%%%%%%%%%%%%%%%%%%%%%%%%%%%%%%%%%%%%%%%%%%%%%%%%%%%%%%%%%%%
\begin{acknowledgments}
B. Pal acknowledges the financial support through an INSPIRE Fellowship 
from the Department of Science and Technology, India. 
\end{acknowledgments}
%%%%%%%%%%%%%%%%%%%%%%%%%%%%%%%%%%%%%%%%%%%%%%%%%%%%%%%%%%%%%%%%%%%%%%%%%%%%

\end{document}